\documentstyle[preprint,eqsecnum,epsf,aps]{revtex}
\begin{document}
\title{Weakly Nonlinear Analysis of Electroconvection \\ in a Suspended Fluid
Film}
\author{V.B. Deyirmenjian, Zahir A. Daya, and Stephen W. Morris}
\address{Department of Physics and Erindale College, \\ University of Toronto,
Toronto, Ontario,
Canada, M5S 1A7}
\date{\today}
\maketitle

\begin{abstract}
It has been experimentally observed that weakly conducting suspended films
of smectic liquid crystals undergo electroconvection when subjected to a
large enough potential difference. The resulting counter-rotating vortices
form a very simple convection pattern and exhibit a variety of
interesting nonlinear effects.  The linear stability problem for this
system has recently been solved. The convection mechanism, which involves
charge separation at the free surfaces of the film, is applicable to any
sufficiently two-dimensional fluid.  In this paper, we derive an amplitude
equation which describes the weakly nonlinear regime, by starting from the
basic electrohydrodynamic equations. This regime has been the subject of
several recent experimental studies.  The lowest order amplitude equation
we derive is of the Ginzburg-Landau form, and describes a forward
bifurcation as is observed experimentally. The coefficients of the
amplitude equation are calculated and compared with the values independently
deduced from the linear stability calculation.

\end{abstract}
\pacs{47.20.K,47.65.+a,61.30.-v}
\narrowtext

\section{Introduction}
\label{intro}

Although spatio-temporal pattern formation is ubiquitous in nature, there are
relatively few systems which are amenable to both accurate experimental study
and first-principles weakly nonlinear analysis.\cite{review} The classic
examples involving fluid mechanical instabilities are Rayleigh-B\'enard
convection and Taylor-Couette flow. The results of perturbation theory based on
the Navier-Stokes and heat equations are in good agreement with high precision
experiments in the weakly nonlinear regime of these two
instabilities.\cite{review,ahlersreview}  A more complex example is
electroconvection in nematic liquid crystals due to the Carr-Helfrich
mechanism.\cite{nematicreview}  Here, remarkably good agreement has been
achieved in spite of the complexity of the problem. However, in other
cases either the materials cannot be sufficiently characterized or the
underlying physical equations are not understood well enough to allow
quantitative comparisons between observations and theory.

\begin{figure}
\epsfxsize =5in
\centerline{\epsffile{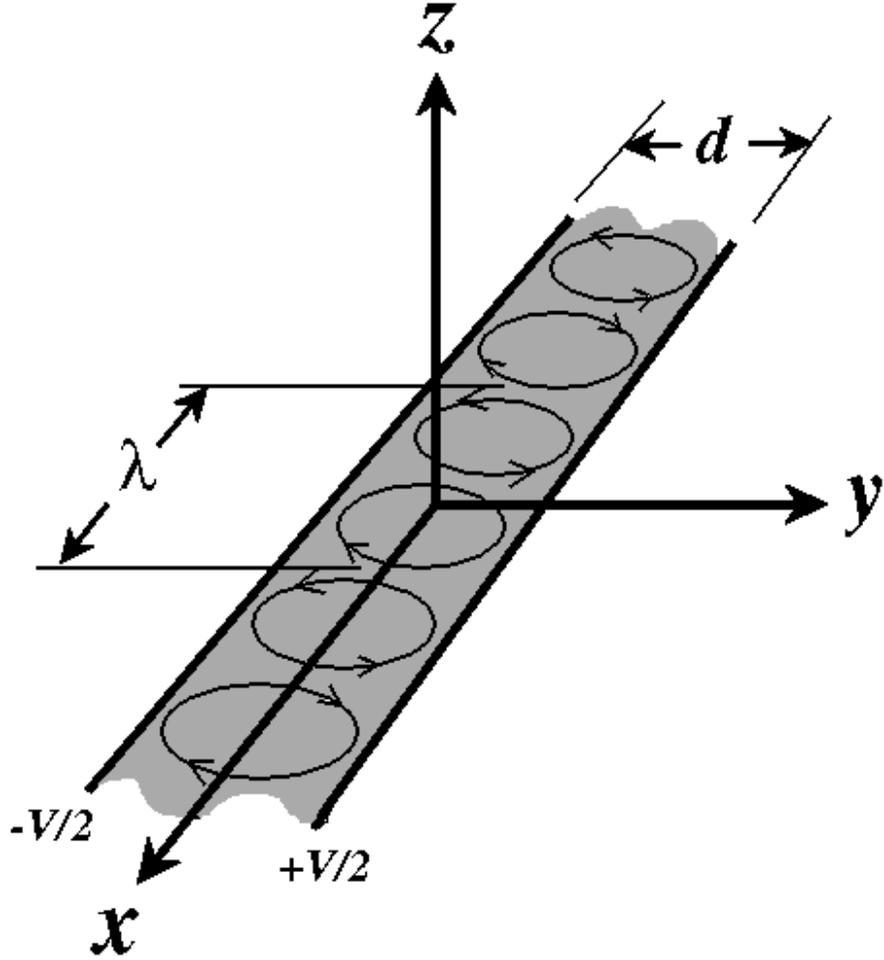}}
\vskip 0.1in
\caption{Schematic of film geometry and coordinate system. The wire electrode
configuration is shown. The vortex pair periodicity is $\lambda = 2 \pi d /
\kappa$, where $d$ is the film width.  The thickness $s$ of the film (not
shown) is such that $s \ll d$.  }
\label{schematic}
\end{figure}

Electroconvection in suspended smectic films is a promising new experimental
system for quantitative studies of spatio-temporal pattern
formation.\cite{Morris,MorrisPRA,MDDMend,MDMgle} When a thin, suspended film of
smectic liquid crystal is subjected to a sufficient potential difference, a
charge separation arises which drives the film into convection. The flow
pattern just above onset is sustained by the electric field acting on charges
which develop near the free surfaces of the film. These charges are simply a
consequence of the electrostatic boundary conditions which must be satisfied by
the
fields inside and outside of the film.\cite{linear} Fig.~\ref{schematic} shows
a
schematic of the experimental arrangement. This source of charge is distinct
from that due to the Carr-Helfrich mechanism which drives bulk
electroconvection in certain nematics.\cite{nematicreview} In that case, the
charge generation mechanism involves an essential coupling to the director
orientation. In experiments on smectic~A liquid crystal
films\cite{Morris,MorrisPRA,MDDMend,MDMgle}, in which the director was
perpendicular to the film, no orientational effects were observed, indicating
that the flow remained isotropic in the film plane. Recent experiments on
smectic C films\cite{paderborn} showed convection and flow alignment of the
projection of the director in the plane of the film, but were not consistent
with the Carr-Helfrich mechanism. These were likely driven by the mechanism
discussed here, with the flow alignment a secondary effect.  It has, however,
been hypothesized\cite{ried} that the two mechanisms might coexist in some
smectic C materials.  The mechanism we discuss here is presumably also
responsible for convection observed in thin, suspended films of isotropic
fluids and in nematics in certain regimes\cite{faetti}. These cases involve
substantial three-dimensional effects, however, because they lack the smectic
layering which has the effect of restricting the flow to the film plane. In
what follows, we consider only very thin isotropic films, relevant to the case
of smectic~A, on which most of the experiments have been
performed.\cite{Morris,MorrisPRA,MDDMend,MDMgle}

A theoretical model of the onset of electroconvection in suspended films was
introduced by Daya, Morris, and de Bruyn.\cite{linear} The film was represented
as a weakly conducting, two-dimensional, isotropic fluid. To find the
electric fields and charge densities which drive convection, the electrostatic
potential was determined. The
three-dimensional electrostatic equations effectively constitute a nonlocal
coupling between the in-plane fields and charge densities which appear in the
two-dimensional Navier-Stokes and charge continuity equations. This extra
coupling formally distinguishes the resulting equations from those of thermal
convection in the Rayleigh-B\'enard problem, although some interesting
similarities remain. The value of the critical wavenumber from the linear
stability analysis\cite{linear} is in good agreement with
experiment\cite{Morris,MorrisPRA,MDMgle}.

The purpose of this paper is to present a weakly nonlinear analysis
of the electrohydrodynamic equations given in Ref.~\cite{linear}. The
multiple-scales perturbation theory employed in our treatment is similar
to that given in Ref.~\cite{review} for Rayleigh-B\'enard convection.
Although there are important physical differences between these two pattern
forming instabilities, the resulting amplitude equation for both problems is of
the Ginzburg-Landau form
\begin{equation}
\tau_0 \partial_t A = \epsilon A + \xi_0^2 \partial_x^2 A -
   g_0 A|A|^2 \;, \label{ampintro}
\end{equation}
where $\epsilon$ is the control parameter, which depends on the applied
electric potential, and $A(x,t)$ is a slowly varying amplitude.
The coefficients $\tau_0$, $\xi_0$, and $g_0$ are compared with those
previously obtained by other methods. In particular, $\tau_0$ and $\xi_0$ are
found to be in good agreement with the values determined from the linear
stability analysis of Daya {\it et al}.\cite{linear} Mao, de Bruyn, and Morris
have experimentally measured all three coefficients.\cite{MDMgle}  The
experimental value of $\xi_0$ is in reasonable agreement with theory.
Quantitative comparison of  $\tau_0$ and $g_0$ with theory is difficult at the
present time due to large uncertainties in the conductivity and viscosity of
the liquid crystal, which are required to non-dimensionalize the experimental
results.

Determining the amplitude equation constitutes a first step towards
understanding the fully nonlinear regime beyond the onset of
electroconvection. For small wavenumbers near threshold,
the stability of solutions of Eqn.~\ref{ampintro} determines the regions
of control parameter-wavenumber space where the vortex pattern itself is
stable.\cite{review} For example, one expects such a one-dimensional pattern to
exhibit a long-wavelength instability due to the Eckhaus
mechanism\cite{review,mannevile} which restricts the range of stable
wavenumbers available to the pattern. The amplitude equation can also be used
to study how the ends of a finite-length film affect the range of stable
wavenumbers. This wavenumber selection mechanism was first investigated by
Cross, Daniels, Hohenberg, and Siggia~\cite{CDHS} in the context of
Rayleigh-B\'enard convection in finite containers. End-selection was observed
experimentally by Mao, Morris, Daya, and de~Bruyn~\cite{MDDMend} in
electroconvection patterns in smectic~A films. It is possible to extend the
present theory to determine the Eckhaus and end-selection stability
boundaries\cite{sidewalltheory}, but this is beyond the scope of the present
paper.

This paper is organized as follows. The linear stability analysis of the
electroconvective instability is briefly reviewed in Section~\ref{linstab}.
The amplitude equation is determined in Section~\ref{derivation}. Section
{}~\ref{discussion} compares the results of this theoretical investigation with
previously obtained observations, and contains a brief conclusion.

\section{Linear Stability Analysis}
\label{linstab}

In this section, the physical model describing electroconvection in
a thin film is presented. The linear stability analysis of the
relevant equations is concisely reviewed. Further details are given in
Ref.~\cite{linear}. Note that we change some of the notation of
Ref.~\cite{linear} to simplify the presentation of this paper.

The film is treated as a two-dimensional conducting fluid in the
$xy$-plane, with areal material parameters $\rho_s = s \rho$,
$\eta_s = s \eta$, and $\sigma_s = s \sigma$, where $s$, $\rho$, $\eta$,
and $\sigma$ are the film thickness, bulk density, bulk molecular
viscosity, and bulk conductivity, respectively. The coordinate system is
shown in Fig.~\ref{schematic}. The film is assumed to be infinite in the
$x$-direction and bounded between $-d/2$ and $d/2$ in the $y$-direction. We
only consider the thin film limit where $s/d \rightarrow 0$. Two electrode
configurations are analyzed. In the `plate' geometry, the film is suspended
between two thin sheet electrodes which fill the rest of the $xy$ plane,
whereas in the `wire' geometry, the film is suspended between two thin line
electrodes which are along the $x$ direction. In both cases, the electrode at
$y=-d/2$ is fixed
at a potential of $-V/2$, and the electrode at $y=d/2$ is at a potential of
$V/2$.

The Navier-Stokes equation

\begin{equation}
\rho_s \biggl[\frac{\partial {\bf u}}{\partial t} + ({\bf
u} \cdot \nabla_s) {\bf u}\biggr]  =  -\nabla_s P_s + \eta_s
{\nabla_s} ^{2} {\bf u} + q {\bf E}_s,
\label{navierstokes}
\end{equation}
describes the fluid flow of the liquid crystal, where
$\nabla_s=(\partial_x,\partial_y,0)$, $P_s(x,y)$, $q(x,y)$, and
${\bf E}_s(x,y)$ are the two-dimensional gradient operator,
two-dimensional pressure field, surface charge density, and electric
field in the film plane, respectively. The incompressibility of the fluid
implies that

\begin{equation}
\partial_x u + \partial_y v = 0 \;, \label{incomp}
\end{equation}
where $u(x,y)$ and $v(x,y)$ are the $x$ and $y$ components of the
two-dimensional velocity field ${\bf u}$. The pressure field is eliminated from
Eqn.~\ref{navierstokes} by applying the curl operator. Taking the curl of
Eqn.~\ref{navierstokes} twice, using
Eqn.~\ref{incomp}, and selecting the $y$ component gives:

\begin{eqnarray}
-\rho_s \partial_t \nabla_s^{2} v + \rho_s \partial_y(\nabla_s
	\cdot [ ({\bf u} \cdot \nabla_s){\bf u} ] ) \nonumber
	-\rho_s \nabla_s^{2} [ ({\bf u} \cdot \nabla_s)v ] & = & \nonumber \\
 -\eta_s \nabla_s^4 v +(\partial_x^2 q)(\partial_y \Psi |_{z=0})
+(\partial_x q)(\partial_{xy} \Psi |_{z=0})
-(\partial_{xy} q)(\partial_x \Psi |_{z=0})
-(\partial_y q)(\partial_x^2 \Psi |_{z=0}) & & , \label{curlnavier}
\end{eqnarray}
where the electric potential $\Psi(x,y,z)$ is related to the in-plane
electric field via ${\bf E}_s(x,y)= - \nabla_s \Psi(x,y,z) |_{z=0}$. The
three-dimensional Laplace equation,
\begin{equation}
\nabla^2 \Psi = 0,
\label{3dlaplace}
\end{equation}
specifies $\Psi$ in the half space $z \geq 0$ with appropriate boundary
conditions in the $xy$-plane, where
$\nabla=(\partial_x,\partial_y,\partial_z)$.
The surface charge density depends on the discontinuity in the
$z-$derivative of $\Psi$ across the two surfaces of the film:

\begin{eqnarray}
q & = & - {\epsilon_0} \frac{\partial \Psi}{\partial z}{\Bigg|_{z=0^+}}
 + {\epsilon_0} \frac{\partial \Psi}{\partial z}{\Bigg|_{z=0^-}}
   \;, \nonumber \\
& = & -2 {\epsilon_0} \frac{\partial \Psi}{\partial z}{\Bigg|_{z=0^+}},
\label{qdefine}
\end{eqnarray}
where $\epsilon_0$ is the permittivity of free space.

The motion of charge is governed by the charge continuity equation
\begin{equation}
\frac{\partial q}{\partial t} =
-\nabla_s \cdot (q{\bf u} + \sigma_s {{\bf E}_s}),
\label{chgcontinuity}
\end{equation}
which includes contributions from both the conductive
$\sigma_s {{\bf E}_s}(x,y)$ and convective $q(x,y){\bf u}(x,y)$
current densities. Diffusion of charge in the plane of the film can be
neglected.

Equations~\ref{curlnavier}-\ref{chgcontinuity} constitute the basic
electrohydrodynamic equations; the electrode geometry enters into the
boundary conditions on $\Psi$. The solution of these equations in the ``base
state"
below the onset of convection has ${\bf u}^{(0)}(x,y)=0$, with $q^{(0)}(y)$ and
$\Psi^{(0)}(y,z)$ satisfying the electrostatic boundary value problem given by
Eqns.~\ref{3dlaplace} and \ref{qdefine}. To examine the stability of the
base state, we introduce the perturbed quantities

\begin{eqnarray}
{\bf u} & = & 0 + {\bf u}^{(1)}, \\
q & = & q^{(0)} + q^{(1)}, \\
{\bf E}_s& = & {\bf E}_s^{(0)} + {\bf E}_s^{(1)},
\label{perturbs}
\end{eqnarray}
where ${\bf E}_s^{(0)} = {E_y}^{(0)}{\bf \hat y}$
and ${\bf E}_s^{(1)}(x,y)={E_{x}}^{(1)}(x,y){\bf \hat x}+
{E_{y}}^{(1)}(x,y){\bf \hat y}$.
Quantities which have dimensions of length, time, charge density, and
electric potential are nondimensionalized by $d$, $\epsilon_{0}d/\sigma_s$,
$\epsilon_{0}V/d$, and $V$, respectively. Substituting the perturbed field
variables into Eqns.~\ref{curlnavier}-\ref{chgcontinuity}, nondimensionalizing,
and suppressing the superscripts, yields

\begin{eqnarray}
\nabla_s^4 v  -{\cal R}\partial_x^2 q +{\cal R}Q\partial_x^{2} \Psi |_{z=0}
   & = & {\cal R} \partial_x [ (\partial_x q)(\partial_y \Psi |_{z=0}) -
    (\partial_y q)(\partial_x \Psi |_{z=0}) ]  \nonumber \\
 +{\cal P}^{-1} \{ \partial_t (\nabla_s^2 v)
 -\partial_y( \nabla_s \cdot [ ({\bf u} \cdot \nabla_s) {\bf u} ])
 +\nabla_s^2[ ({\bf u} \cdot \nabla_s)v ] \} ,\label{directeq1} \\
-Q v +\nabla_s^{2} \Psi |_{z=0}   & = &
     \partial_t q + u(\partial_x q) + v(\partial_y q) , \label{directeq2}  \\
     q + (\partial_z \Psi)|_{z=0^{+}} - (\partial_z \Psi)|_{z=0^{-}}
     & = & 0 , \label{directeq3} \\
     \nabla^{2} \Psi & = & 0 \label{directeq4} \;.
\end{eqnarray}
The dimensionless parameters ${\cal R}$ and ${\cal P}$ are analogous to
the Rayleigh and Prandtl numbers. We will henceforth consider only the limit
${\cal P} \rightarrow \infty$, as this is the case most relevant to experiments
on real smectic materials\cite{MDMgle}, for which ${{\cal P} \approx 10-100}$.
The non-constant coefficient $Q(y)$ depends on the electrode configuration and
is given by $Q(y)=\partial_y q^{(0)}(y)$, where $q^{(0)}(y)$ is the base state
charge density. The variables $v$, $q$, and $\Psi$ above represent the
dimensionless perturbed functions $v^{(1)}$, $q^{(1)}$, and
$\Psi^{(1)}$, respectively, and satisfy the following boundary conditions:

\begin{equation}
v(x,y=\pm \frac{1}{2})=(\partial_y v)(x,y=\pm \frac{1}{2})=0  \;,
\label{directbc1}
\end{equation}
\begin{equation}
\Psi(x,y=\pm \frac{1}{2},0) = 0 \;,
\label{directbc2}
\end{equation}
\begin{equation}
\Psi(x,y,z) \rightarrow 0 \;, z \rightarrow \pm \infty \;.
\label{directbc3}
\end{equation}
In the plate electrode configuration, Dirichlet boundary conditions are
employed on the $xy$ plane, with
\begin{equation}
\Psi(x,y,0) = 0, {~~~}  |y| > \frac{1}{2} .
\label{directbc4}
\end{equation}
In the wire electrode geometry, mixed boundary conditions apply such that
\begin{equation}
\partial_z{\Psi(x,y,z)} |_{z=0^+} = 0, {~~~} |y| > \frac{1}{2} \;.
\label{directbc5}
\end{equation}
In both cases the potential $\Psi(x,y,0) =\Psi |_{z=0}(x,y)$, $|y| \leq 0$,
is specified on the film.

Equations \ref{directeq1}-\ref{directeq4} can be expressed as

\begin{equation}
{\cal L}{\cal C} = {\cal B} \;, \label{diropeqn}
\end{equation}
where

\begin{equation}
{\cal L} = \left(
\begin{array}{cccc}
\nabla_s^4 & -{\cal R}\partial_x^2 & {\cal R}Q\partial_x^{2} & 0 \\
-Q         & 0                     & \nabla_s^2              & 0   \\
 0 & 1 & 0 & \partial_z(.)|_{z=0^+} - \partial_z(.)|_{z=0^-}  \\
 0   & 0 & 0 & \nabla^2  \\
\end{array}\label{directop}
\right),
\end{equation}

\begin{equation}
{\cal C} = \left(
\begin{array}{c}
v(x,y) \\
q(x,y) \\
\Psi(x,y,z) |_{z=0} \\
\Psi(x,y,z) \\
\end{array}
\right),  \label{directC}
\end{equation}
and

\begin{equation}
{\cal B} = \left(
\begin{array}{c}

{\cal R} \partial_x [ (\partial_x q)(\partial_y \Psi |_{z=0}) -
     (\partial_y q)(\partial_x \Psi |_{z=0}) ] \\
\partial_t q + u(\partial_x q) + v(\partial_y q) \\
0 \\
0 \\
\end{array}
\right).
\end{equation}
The linear stability problem is defined by

\begin{equation}
{\cal L}{\cal C} = 0 \;. \label{linearstab}
\end{equation}
The neutral
stability curve ${\cal R} = {\cal R}_c(\kappa)$ is determined by
substituting the normal mode solution

\begin{equation}
{\cal C} = \left(
\begin{array}{c}
\bar{v}_{\kappa}(y)  \\
\bar{q}_{\kappa}(y)  \\
\bar{\Psi}_{\kappa}(y,0)  \\
\bar{\Psi}_{\kappa}(y,z)  \\
\end{array}
\right) e^{i \kappa x } = \bar{{\cal C}}_{\kappa} e^{i \kappa x} \;,
\label{normalmode}
\end{equation}
into Eqn.~\ref{linearstab}. The following alterations have been
made to the notation of Ref.~\cite{linear}:
$\Lambda(y) \rightarrow \bar{v}_{\kappa}(y)$,
$\Theta(y) \rightarrow \bar{q}_{\kappa}(y)$,
$\Omega_{s}(y) \rightarrow \bar{\Psi}_{\kappa}(y,0)$, and
$\Omega(y,z) \rightarrow \bar{\Psi}_{\kappa}(y,z)$.  The variables
$\bar{v}_{\kappa}$, $\bar{q}_{\kappa}$, and $\bar{\Psi}_{\kappa}$ are
expanded as

\begin{eqnarray}
\bar{v}_{\kappa}(y) & = & \sum_{m=1}^{\infty} \bar{A}_m
       \bar{v}_{\kappa m}(y) \;, \label{directsumsol1} \\
\bar{q}_{\kappa}(y) & = & \sum_{m=1}^{\infty} \bar{A}_m
       \bar{q}_{\kappa m}(y) \;, \label{directsumsol2} \\
\bar{\Psi}_{\kappa}(y,z) & = &
      \sum_{m=1}^{\infty} \bar{A}_m \bar{\Psi}_{\kappa m}(y,z) \;,
  \label{directsumsol3}
\end{eqnarray}
where $\bar{v}_{\kappa m}(y)$, $\bar{q}_{\kappa m}(y)$, and
$\bar{\Psi}_{\kappa m}(y,z)$ satisfy the boundary conditions
Eqns.~\ref{directbc1} to~\ref{directbc5}. The linear problem is solved
numerically in Ref.~\cite{linear} by substituting $C_m$, the even Chandrasekhar
function\cite{Chandrasekhar}, for
$\bar{v}_{\kappa m}(y)$ and then finding self-consistent solutions for
$\bar{q}_{\kappa m}(y)$ and  $\bar{\Psi}_{\kappa m}(y,z)$.

\section{Derivation of the Amplitude Equation}
\label{derivation}

The multiple-scales approach is used to obtain the
amplitude equation, which describes the slow temporal and spatial variation of
the field variables.\cite{review,mannevile} The slow scales
$X=\epsilon^{1/2}x$ and $T=\epsilon t$ are treated as independent of the fast
scales $x$ and $t$. We choose $\epsilon=({\cal R}-{\cal
R}_{c0})/{\cal R}_{c0}$, where ${\cal R}_{c0}$ is the critical value of
${\cal R}$ at the minimum of the neutral stability curve
${\cal R}={\cal R}_c(\kappa)$. The nonlinear equation describing the system,
Eqn.~\ref{diropeqn}, is expanded in powers of $\epsilon^{1/2}$ as
follows:
\begin{eqnarray}
{\cal L} & = & {\cal L}_0 + \epsilon^{1/2}{\cal L}_1 + \epsilon{\cal L}_2
  + ... \;, \\
{\cal C} & = & \epsilon^{1/2}{\cal C}_0 + \epsilon{\cal C}_1
  + \epsilon^{3/2}{\cal C}_2 +... \;, \\
{\cal B} & = & \epsilon^{1/2}{\cal B}_0 + \epsilon{\cal B}_1
  + \epsilon^{3/2}{\cal B}_2 + ... \;,
\end{eqnarray}
and
\begin{eqnarray}
v(x,y)& = & \epsilon^{1/2}v_0(x,y) + \epsilon v_1(x,y)
   +\epsilon^{3/2}v_2(x,y) +... \;, \\
q(x,y)& = & \epsilon^{1/2}q_0(x,y) + \epsilon q_1(x,y)
   +\epsilon^{3/2}q_2(x,y) +... \;, \\
\Psi(x,y,z)& = & \epsilon^{1/2}\Psi_0(x,y,z) + \epsilon\Psi_1(x,y,z)
   +\epsilon^{3/2}\Psi_2(x,y,z) + ... \;.
\end{eqnarray}
The differentials $\partial_x$ and $\partial_t$ in the original equations
transform as
$\partial_x \rightarrow \partial_x + \epsilon^{1/2}\partial_X$ and
$\partial_t \rightarrow \partial_t + \epsilon\partial_T$.

At orders $\epsilon^{1/2}$, $\epsilon$, and $\epsilon^{3/2}$,
Eqn.~\ref{diropeqn} becomes
\begin{eqnarray}
{\cal L}_0 {\cal C}_0 & = & {\cal B}_0  \;,\label{orderhalf} \\
{\cal L}_0{\cal C}_1 + {\cal L}_1{\cal C}_0 & = & {\cal B}_1 \;,
\label{orderone} \\
{\cal L}_0{\cal C}_2 + {\cal L}_1{\cal C}_1 + {\cal L}_2{\cal C}_0
   & = & {\cal B}_2  \label{orderthreehalves} \;,
\end{eqnarray}
respectively.

The most general solution of
Eqn.~\ref{orderhalf}, at order $\epsilon^{1/2}$, is
\begin{equation}
{\cal C}_0 = \left(
\begin{array}{c}
v_0(x,y) \\
q_0(x,y) \\
\Psi_0(x,y,z)|_{z=0} \\
\Psi_0(x,y,z) \\
\end{array}
\right)
=A_0(X,T) \left(
\begin{array}{c}
\bar{v}_0(y) \\
\bar{q}_0(y) \\
\bar{\Psi}_0(y,0) \\
\bar{\Psi}_0(y,z) \\
\end{array}
\right)e^{i\kappa_0 x} + c.c. \;, \label{orderhalfsol}
\end{equation}
where $A_0(X,T)$ is the amplitude function, $\kappa_0$ is the critical
wavenumber
which minimizes the function ${\cal R}_c(\kappa)$, and $c.c.$ denotes
complex conjugation. The functions
$\bar{v}_0(y)=\bar{v}_{\kappa}(y)|_{\kappa=\kappa_0}$,
$\bar{q}_0(y)=\bar{q}_{\kappa}(y)|_{\kappa=\kappa_0}$, and
$\bar{\Psi}_0(y,z)=\bar{\Psi}_{\kappa}(y,z)|_{\kappa=\kappa_0}$
are solutions of the linear stability problem, Eqns.~\ref{linearstab},
\ref{normalmode}-\ref{directsumsol3}.

To solve the order $\epsilon$ equation, Eqn.~\ref{orderone},
the relation

\begin{equation}
\frac{\partial}{\partial \kappa} \left[ {\cal L} \left(
\begin{array}{c}
\bar{v}_{\kappa}(y) e^{i \kappa x} \\
\bar{q}_{\kappa}(y) e^{i\kappa x} \\
\bar{\Psi}_{\kappa}(y,0) e^{i\kappa x} \\
\bar{\Psi}_{\kappa}(y,z) e^{i\kappa x} \\
\end{array}
\right) \right]_{\kappa=\kappa_0} = 0 \;,  \label{transformeq}
\end{equation}
is employed. This is used to transform Eqn.~\ref{orderone} to

\begin{equation}
{\cal L}_0{\tilde {\cal C}}_1 = {\cal B}_1 \;, \label{eqtilde}
\end{equation}
which is solved by the ansatz

\begin{eqnarray}
{\tilde {\cal C}}_1 & \equiv & \left(
\begin{array}{c}
{\tilde v}_1(x,y) \\
{\tilde q}_1(x,y) \\
{\tilde \Psi}_1(x,y,0) \\
{\tilde \Psi}_1(x,y,z) \\
\end{array}
\right)  \nonumber \\
& = & \left\{ \left(
\begin{array}{c}
v^{\epsilon}_1(y) \\
q^{\epsilon}_1(y) \\
{\Psi}^{\epsilon}_1(y,0) \\
{\Psi}^{\epsilon}_1(y,z) \\
\end{array}
\right) A_0^2 e^{2i\kappa_0 x} + \left(
\begin{array}{c}
\bar{v}_0(y) \\
\bar{q}_0(y) \\
\bar{\Psi}_0(y,0) \\
\bar{\Psi}_0(y,z) \\
\end{array}
\right) A_1 e^{i\kappa_0 x} + c.c. \right\} + \left(
\begin{array}{c}
v^{\epsilon}_2(y) \\
q^{\epsilon}_2(y) \\
{\Psi}^{\epsilon}_2(y,0) \\
{\Psi}^{\epsilon}_2(y,z) \\
\end{array}
\right) |A_0|^2 \;. \label{trialsolution}
\end{eqnarray}
The variable $A_1(X,T)$ is a second amplitude function. Substitution
of Eqn.~\ref{trialsolution} into Eqn.~\ref{eqtilde} gives
the following sets of partial differential equations

\begin{eqnarray}
(\partial_y^2 - (2\kappa_0)^2)^2 v^{\epsilon}_1 + (2\kappa_0)^2 {\cal R}_{c0}
q^{\epsilon}_1
  - (2\kappa_0)^2{\cal R}_{c0}Q {\Psi}^{\epsilon}_1|_{z=0}  & = &
  -2\kappa_0^2 {\cal R}_{c0} [ \bar{q}_0(\partial_y \bar{\Psi}_0 |_{z=0})
  -(\partial_y \bar{q}_0)\bar{\Psi}_0 |_{z=0} ] \;, \label{I1} \\
-Qv^{\epsilon}_1 + (\partial_y^2-(2\kappa_0)^2) {\Psi}^{\epsilon}_1 |_{z=0}  &
= &
  i\kappa_0 \bar{u}_0 \bar{q}_0 + \bar{v}_0(\partial_y \bar{q}_0) \;, \label{I2} \\
       q^{\epsilon}_1 + (\partial_z {\Psi}^{\epsilon}_1)|_{z=0^+} -(\partial_z
{\Psi}^{\epsilon}_1)|_{z=0^-} & = & 0 \;,
	\label{I3} \\
     (\partial_y^2 + \partial_z^2 - (2\kappa_0)^2) {\Psi}^{\epsilon}_1 & = & 0
\;, \label{I4}
\end{eqnarray}
and

\begin{eqnarray}
\partial_y^4 v^{\epsilon}_2  & = & 0 \;, \label{II1} \\
-Qv^{\epsilon}_2 + \partial_y^2 {\Psi}^{\epsilon}_2 |_{z=0} & = &
  -i\kappa_0 \bar{u}_0 \bar{q}_0^* + i\kappa_0 \bar{u}_0^* \bar{q}_0
  +\bar{v}_0(\partial_y \bar{q}_0^*) +\bar{v}_0^*(\partial_y \bar{q}_0)
	\;, \label{II2} \\
       q^{\epsilon}_2 + (\partial_z {\Psi}^{\epsilon}_2)|_{z=0^+} - (\partial_z
{\Psi}^{\epsilon}_2)|_{z=0^-}
      & = & 0 \;, \label{II3} \\
     (\partial_y^2 + \partial_z^2) {\Psi}^{\epsilon}_2 & = & 0 \;, \label{II4}
\end{eqnarray}
where the superscript $^*$ denotes complex conjugation. A vector
${\tilde {\cal C}}_1$ which solves Eqn.~\ref{eqtilde} is presented in Appendix
A.
The general solution at
order $\epsilon$ is

\begin{equation}
{\cal C}_1 = {\tilde {\cal C}}_1 -
      \{ (2\kappa_0)^{-1}(2\partial_x\partial_X A_0)e^{i\kappa_0 x}
      {\bar {\cal C}}_0' + c.c. \} \;,
\end{equation}
where the prime denotes $\partial_{\kappa}$, and ${\bar {\cal C}}_0'$ is given
by
\begin{equation}
{\bar {\cal C}}_0' = \left(
\begin{array}{c}
\bar{v}_0'(y) \\
\bar{q}_0'(y) \\
\bar{\Psi}_0'(y,0) \\
\bar{\Psi}_0'(y,z) \\
\end{array}
\right) = \left(
\begin{array}{c}
(\partial_{\kappa}\bar{v}_{\kappa}(y)) |_{\kappa=\kappa_0} \\
(\partial_{\kappa}\bar{q}_{\kappa}(y)) |_{\kappa=\kappa_0} \\
(\partial_{\kappa}\bar{\Psi}_{\kappa}(y,0)) |_{\kappa=\kappa_0} \\
(\partial_{\kappa}\bar{\Psi}_{\kappa}(y,z)) |_{\kappa=\kappa_0} \\
\end{array}
\right) \;.
\end{equation}

The order $\epsilon^{3/2}$ equation, Eqn.~\ref{orderthreehalves}, can similarly
be transformed to
\begin{equation}
{\cal L}_0 {\tilde {\cal C}}_2 = {\cal G} \;, \label{nonhom}
\end{equation}
by using Eqn.~\ref{transformeq}. To establish the condition for the existence
of a solution
${\tilde {\cal C}}_2$ of Eqn.~\ref{nonhom}, the inner product
\begin{eqnarray}
({\cal C}_b,{\cal C}_a) & = & (2\pi / \kappa_0)^{-1}
   \int_0^{2\pi / \kappa_0} dx
   \int_{-\infty}^{\infty} dy \;
  \{ v_b^*(x,y)v_a(x,y) +q_b^*(x,y)q_a(x,y) +
\Psi_b^* |_{z=0}\Psi_a |_{z=0} \} \nonumber \\
 & & + (2\pi / \kappa_0)^{-1} \int_0^{2\pi / \kappa_0} dx
\int_{-\infty}^{\infty} dy \int_{-\infty}^{\infty} dz \;
\{ \Psi_b^*(x,y,z) \Psi_a(x,y,z) \} \;, \label{innerproduct}
\end{eqnarray}
is employed, where ${\cal C}_i$ ($i=a,b$) is
\begin{equation}
{\cal C}_i = \left(
\begin{array}{c}
v_i(x,y) \\
q_i(x,y) \\
\Psi_i(x,y,z) |_{z=0} \\
\Psi_i(x,y,z) \\
\end{array}
\right)~.
\end{equation}
The adjoint operator ${\cal L}_0^{\dag}$ is determined by integrating
$({\cal C}_b,{\cal L}_0{\cal C}_a)$ by parts:
\begin{eqnarray}
({\cal C}_b,{\cal L}_0{\cal C}_a) & = &
   ({\cal L}_0^{\dag}{\cal C}_b,{\cal C}_a) + {\hbox{\rm boundary terms.}}
  \label{intbyparts1}
\end{eqnarray}
The homogeneous equation ${\cal L}_0{\cal C}_a=0$ with homogeneous boundary
conditions, Eqns.~\ref{directbc1}-\ref{directbc5}, will be referred to as the
homogeneous ``direct" problem. For this case, the left-hand side of
Eqn.~\ref{intbyparts1} is zero. Some of the boundary terms on the
right-hand side of Eqn.~\ref{intbyparts1} vanish because
of the boundary conditions on the direct variables
${\cal C}_a$. By defining homogeneous boundary conditions for the adjoint
quantities ${\cal C}_b$, the remaining boundary terms can be set to zero.
This implies that

\begin{equation}
{\cal L}_0^{\dag}{\cal C}_b=0 \;, \label{adjequation}
\end{equation}
or more explicitly

\begin{eqnarray}
\nabla_s^4 v_b -Q q_b    & = & 0 \label{adjequation1} \;, \\
-{\cal R}_{c0} \partial_x^2 v_b + \Psi_b |_{z=0}
    & = & 0 \label{adjequation2} \;, \\
{\cal R}_{c0} Q \partial_x^2 v_b  + \nabla_s^2 q_b
  +(\partial_z \Psi_b)|_{z=0^+} - (\partial_z \Psi_b)|_{z=0^-}
     & = & 0 \;, \label{adjequation3} \\
 \nabla^2 \Psi_b & = & 0 \;. \label{adjequation4}
\end{eqnarray}
Since ${\cal L}_0 \neq {\cal L}_0^{\dag}$, $ {\cal L}_0$ is not self-adjoint.
The corresponding operator in the Rayleigh-B\'enard problem can
be made self-adjoint by an appropriate choice of the inner
product.\cite{review}
This does not appear to be possible in the present case. The boundary
conditions on the adjoint quantities are
\begin{equation}
v_b(x,y=\pm \frac{1}{2})=(\partial_y v_b)(x,y=\pm \frac{1}{2})=0  \;,
\label{adjbc1}
\end{equation}
\begin{equation}
\Psi_b(x,y=\pm \frac{1}{2},0) = q_b(x,y=\pm \frac{1}{2}) = 0 \;,
\label{adjbc2}
\end{equation}
\begin{equation}
\Psi_b \rightarrow 0 \;, z \rightarrow \pm \infty \;.
\label{adjbc3}
\end{equation}
As in the direct problem, Dirichlet boundary conditions are employed on the
$xy$ plane in the plate electrode configuration with
\begin{equation}
\Psi_b(x,y,0) = 0, {~~~}  |y| > \frac{1}{2} ,
\label{adjbc4}
\end{equation}
and mixed boundary conditions are applied in the wire electrode geometry with
\begin{equation}
\partial_z{\Psi_b(x,y,z)} |_{z=0^+} = 0, {~~~} |y| > \frac{1}{2} .
\label{adjbc5}
\end{equation}
In both cases the relation $\Psi_b(x,y,0) =\Psi_b |_{z=0}(x,y)$,
$|y| \leq 0$, is specified on the film.

The adjoint problem defined by Eqn.~\ref{adjequation} with boundary
conditions Eqns.~\ref{adjbc1}-\ref{adjbc5} is satisfied by
\begin{equation}
{\cal C}_b = \left(
\begin{array}{c}
\bar{v}_{b\kappa_0}(y)  \\
\bar{q}_{b\kappa_0}(y)  \\
\bar{\Psi}_{b\kappa_0}(y,0)  \\
\bar{\Psi}_{b\kappa_0}(y,z)  \\
\end{array}
\right) e^{i \kappa_0 x } = \bar{{\cal C}}_{b\kappa_0} e^{i \kappa_0 x} \;.
  \label{adjsol}
\end{equation}
Equations~\ref{adjequation} and \ref{adjsol} are analogous to
Eqns.~\ref{linearstab} and \ref{normalmode} in the linear stability problem.
The details of the adjoint solution, Eqn.~\ref{adjsol}, are given in Appendix
B.

Expanding the product $({\cal C}_b,{\cal L}_0{\cal C}_a)$ for the
inhomogeneous direct equation ${\cal L}_0{\cal C}_a={\cal G}$, with
homogeneous boundary conditions Eqns.~\ref{directbc1}-\ref{directbc5},
leads to

\begin{eqnarray}
({\cal C}_b,{\cal G}) & = &
   ({\cal L}_0^{\dag}{\cal C}_b,{\cal C}_a) + {\hbox{\rm boundary terms.}}
\label{intbyparts2}
\end{eqnarray}
The vector ${\cal C}_b$ is specified by Eqn.~\ref{adjsol}.
The right-hand side of Eqn.~\ref{intbyparts2} is zero due to
Eqn.~\ref{adjequation} and the boundary conditions on the direct and adjoint
quantities. Equation~\ref{intbyparts2} therefore reduces to

\begin{equation}
({\cal C}_b,{\cal G})  =  0 \;, \label{solvability}
\end{equation}
where

\begin{eqnarray}
{\cal G} & = & \left(
\begin{array}{c}
G_1 \\
G_2 \\
G_3 \\
G_4 \\
\end{array}
\right) e^{i\kappa_0 x} + c.c. + \; ...  \;. \label{Gvector}
\end{eqnarray}
According to the Fredholm Theorem, Eqn.~\ref{solvability}
is a necessary and sufficient condition for the existence of a solution of
Eqn.~\ref{nonhom}.\cite{Stakgold}
The remaining terms in Eqn.~\ref{Gvector} do not contribute to
Eqn.~\ref{solvability} since they do not depend on $e^{\pm i\kappa_0 x}$.
Substituting Eqn.~\ref{adjsol} and Eqn.~\ref{Gvector} into
Eqn.~\ref{solvability} yields

\begin{eqnarray}
(\bar{{\cal C}}_{b0}e^{i\kappa_0 x},(G_i)e^{i\kappa_0 x} + c.c.) & = &
 ( \;
(\bar{v}_{b0},\bar{q}_{b0},\bar{\Psi}_{b0}|_{z=0},\bar{\Psi}_{b0})e^{i\kappa_0
x},
   (G_1,G_2,G_3,G_4)e^{i\kappa_0 x} + c.c. \; ) \nonumber \\
 & = & (2\pi / \kappa_0)^{-1} \int_{0}^{2\pi / \kappa_0} dx
\int_{-\infty}^{\infty} dy \;
  \{ \; \bar{v}_{b0}^* G_1 + \bar{q}_{b0}^* G_2 +
    \bar{\Psi}_{b0}^*|_{z=0} G_3 \; \}
  \nonumber \\
 &   & + (2\pi / \kappa_0)^{-1} \int_{0}^{2\pi / \kappa_0} dx
\int_{-\infty}^{\infty} dy
       \int_{-\infty}^{\infty} dz \; \bar{\Psi}_{b0}^* G_4  \;, \label{integral4}
\end{eqnarray}
where
$\bar{{\cal C}}_{b0}=\bar{{\cal C}}_{b\kappa_0}$,
$\bar{v}_{b0}=\bar{v}_{b\kappa_0}$,
$\bar{q}_{b0}=\bar{q}_{b\kappa_0}$, and
$\bar{\Psi}_{b0}=\bar{\Psi}_{b\kappa_0}$. Completely expanding
Eqn.~\ref{integral4} gives the amplitude equation

\begin{equation}
F_1 \partial_T A_0 + F_2 A_0 + F_3 (2i\kappa_0 \partial_X)^2 A_0
   + F_4 A_0|A_0|^2 = 0 \;, \label{amp1}
\end{equation}
in the slow scales $X$ and $T$. The coefficients $F_i$ are

\begin{eqnarray}
F_1 & = & \int_{-1/2}^{1/2} dy \;
   \bar{q}_{b0}^{*} \bar{q}_0 \;, \label{eqC1} \\
F_2 & = & \int_{-1/2}^{1/2} dy \{ \; -\kappa_0^2 {\cal R}_{c0} \bar{v}_{b0}^{*}
    ( \bar{q}_0 - Q\bar{\Psi}_0|_{z=0} ) \; \}
 \;, \label{eqC2} \\
F_3 & = &  \int_{-1/2}^{1/2} dy \{ \; (2\kappa_0)^{-1} \bar{v}_{b0}^{*}
 [ 2(\partial_y^2-\kappa_0^2)\bar{v}_0^{'} - {\cal R}_{c0}\bar{q}_0^{'}
  +{\cal R}_{c0} Q\bar{\Psi}_0^{'}|_{z=0} ]
 - \bar{v}_{b0}^{*} \bar{v}_0
 + (2\kappa_0)^{-1} \bar{q}_{b0}^{*} \bar{\Psi}_0^{'}|_{z=0} \; \} \nonumber \\
    &   & + \int_{-\infty}^{\infty} dy \int_{-\infty}^{\infty} dz \;
   (2\kappa_0)^{-1} \bar{\Psi}_{b0}^{*} \bar{\Psi}_0^{'} \;, \label{eqC3} \\
F_4 & = &  \int_{-1/2}^{1/2} dy \{ \; (i\kappa_0)^2 {\cal R}_{c0}
\bar{v}_{b0}^{*}
   [ -\bar{q}_0^{*}(\partial_y {\Psi}^{\epsilon}_1|_{z=0})
   -2(\partial_y \bar{q}_0^{*} ) {\Psi}^{\epsilon}_1 |_{z=0}
   +2q^{\epsilon}_1(\partial_y \bar{\Psi}_0^{*} |_{z=0}) \nonumber \\
 &  & + (\partial_y q^{\epsilon}_1) \bar{\Psi}_0^{*} |_{z=0} +
     \bar{q}_0(\partial_y {\Psi}^{\epsilon}_2 |_{z=0})
   -(\partial_y q^{\epsilon}_2) \bar{\Psi}_0 |_{z=0} ) ]  \nonumber \\
   &    & +\bar{q}_{b0}^{*}
 [ \frac{1}{2}(\partial_y v^{\epsilon}_1)\bar{q}_0^{*}
   +(2i\kappa_0)\bar{u}_0^{*} q^{\epsilon}_1 +\bar{v}_0^{*}(\partial_y
q^{\epsilon}_1)
   +v^{\epsilon}_1(\partial_y \bar{q}_0^{*}) +\bar{v}_0(\partial_y
q^{\epsilon}_2) ] \; \}  \;, \label{eqC4}
\end{eqnarray}
where the prime denotes $\partial_{\kappa}$.
In terms of the fast variables $x$ and $t$, Eqn.~\ref{amp1} is expressed as
\begin{equation}
\tau_0 \partial_t A = \epsilon A + \xi_0^2 \partial_x^2 A -
   g_0 A|A|^2 \;, \label{amplitude}
\end{equation}
such that $A(x,t)=\epsilon^{1/2}A_0(X,T)$, $\tau_0=-F_1/F_2$,
$\xi_0^2=-4\kappa_0^2 F_3/F_2$, and $g_0=-F_4/F_2$.

The normalization of the amplitude function in the solution of the
$\epsilon^{1/2}$ equation, Eqn.~\ref{orderhalfsol}, is
arbitrary. The scale of $A(x,t)$ can be set by requiring
\begin{equation}
Nu - 1 = \langle qv \rangle / \langle \sigma_s E_y \rangle  = |A|^2 \;,
\label{Nusselt}
\end{equation}
where
\begin{equation}
\langle ... \rangle = (2\pi / \kappa_0)^{-1}  \int_0^{2\pi / \kappa_0} dx
   \int_{-1/2}^{1/2} dy  (...) \;.
\end{equation}
Note that $Nu$ is the ``Nusselt number" for the electroconvection problem,
which is defined to be the ratio of the total current density to the
conducted current density, spatially averaged.

\section{Discussion and Conclusion}
\label{discussion}

To find the coefficients of the amplitude equation, Eqn.~\ref{amplitude},
we evaluate Eqns.~\ref{eqC1}-\ref{eqC4} using the numerical techniques
described in Refs.~\cite{linear} and \cite{recipes}. The $y$-integrations
are performed by the Romberg method. A simple SOR algorithm is employed to
solve the Helmholtz equations, Eqns.~\ref{I4}, \ref{II4}, and
\ref{adjpsiyz0}, on a $N \times N$ grid in the first quadrant of the $yz$
plane. The solutions in the rest of $yz$ plane follow from symmetry.
The double integration in Eqn \ref{eqC3} is performed using a 2d trapezoidal
rule based on the same grid. The coefficients are extrapolated such that
$N \rightarrow \infty$ and $N_{film}/N \rightarrow 0$, where $2N_{film}$ is
the number of grid points across the width of the film. The Fourier series
in Eqns.~\ref{nonhomf3}, \ref{generf6}, and \ref{vbarb0m} are expanded up
to $l=29$. Six modes are employed in the solutions
Eqns.~\ref{directsumsol1}-\ref{directsumsol3}, Eqn.~\ref{expansionf1},
and Eqns.~\ref{adjsumsol1}-\ref{adjsumsol3}, of the linear stability,
order $\epsilon$, and adjoint problems, respectively.
Including more terms in these series expansions does not
significantly change our final results.

The values of $\tau_0$, $\xi_0$, and $g_0$  are shown in
Table~\ref{coefftable}. Only $g_0$, the coefficient of the nonlinear term,
depends on the normalization of $A$ by the Nusselt number according to
Eqn.~\ref{Nusselt}.  These results can be compared with those obtained
independently from the linear
stability
calculations of Daya {\it et al.}\cite{linear} In the latter approach, the
correlation length $\xi_0$ was derived from the curvature of the neutral
curve at $\kappa_0$ and the characteristic time $\tau_0$ from the linear
growth rate at $\kappa_0$. Both $\tau_0$ and $\xi_0$ from the linear
stability analysis are in good agreement with the present calculation. This
provides a useful independent check of our numerical results.

The comparison of our theoretical results with the experiments of Mao
{\it et al.}\cite{MDMgle} is difficult at the present time due to the
uncertainties in the measurements of the material parameters of the liquid
crystal film. For example, to nondimensionalize the experimentally measured
value of $g_0$, the factor $\epsilon_0^2 / \sigma^2 s^2$ must be employed.
While the thickness $s$ of the smectic thin
film can be measured accurately, the bulk conductivity of the liquid crystal is
much less well characterized.  Our calculated values of the (Nusselt
normalized) value of $g_0$ are substantially larger than those estimated from
experiment\cite{MDMgle}, but in view of the uncertainty in $\sigma$ (roughly a
factor of 3), no more precise comparison can currently be made.  Experiments
which will more accurately measure $\sigma$ and the viscosity $\eta$ in annular
films are presently being performed.\cite{dayaunpub}

\begin{figure}
\epsfxsize =8in
\centerline{\epsffile{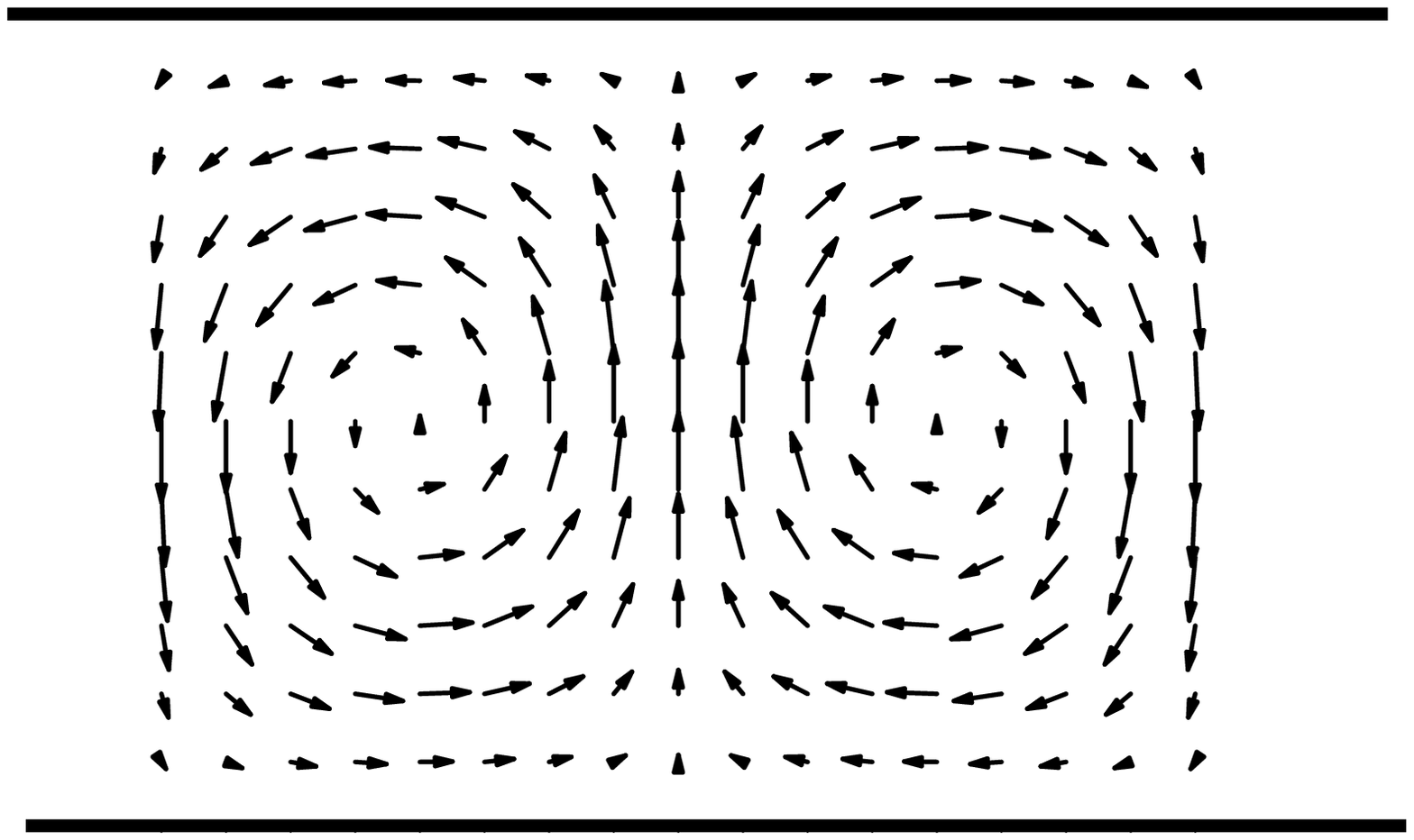}}
\vskip 1.0in
\caption{Vortex pattern just above onset. The dimensionless velocity must be
scaled by  $s \sigma / \epsilon_0$, where $s$, $\sigma$, and $\epsilon_0$
are the film thickness, bulk conductivity, and permittivity of free space,
respectively.  Here we plot the vector velocity field for wire electrodes with
control parameter $\epsilon=0.1$. Using $s=142$~nm and
$\sigma = 2.0 \times 10^{-7}$ ($\Omega$m)$^{-1}$, which are typical values
for smectic films, gives $s \sigma / \epsilon_0 = 3.2$~mm/s.  The
magnitude of the velocity at the centre of the figure is approximately
equal to $2.4$ mm/s. \label{vectorplot}}
\end{figure}

The flow pattern which develops just above onset can be visualized by
evaluating the
velocity vector field ${\bf u}$ on the $xy$ plane. The lowest order $x$ and
$y$ components of the field are given by Eqns.~\ref{unot}
and~\ref{orderhalfsol},
respectively. The amplitude function $A$ in these expressions is obtained
by solving Eqn.~\ref{amplitude} for the steady state case with a specified
control parameter $\epsilon$. An example of the resulting vortex pattern is
shown
in Fig.~\ref{vectorplot}.  This may be qualitatively compared with the
experimental pattern
shown in Fig.~6(b) of Ref.~\cite{MorrisPRA}. As above, a quantitative
comparison of theoretical and experimental velocities is difficult because of
the experimental uncertainty in $\sigma$.

In conclusion, a multiple-scales expansion of the basic electrohydrodynamic
equations for electroconvection in a suspended fluid film was
used to find the lowest order amplitude equation. The set of basic
equations were not self-adjoint, necessitating the evaluation of the adjoint
eigenfunctions. The coefficients $\tau_0$, $\xi_0$, and $g_0$ of the resulting
Ginzburg-Landau equation were determined by
numerical integration. The results of this work can be employed in further
studies of the weakly nonlinear phenomena near the onset of electroconvection
in suspended smectic films. Of particular interest is the mechanism of
wavelength selection\cite{review} and effect of sidewalls on the convection
pattern\cite{MDDMend,sidewalltheory}.

\acknowledgements

We would like to thank John R. de Bruyn for numerous discussions.
This research was supported by the Natural Sciences and Engineering
Research Council of Canada.

\appendix
\section{}

In this appendix, a method for solving the order $\epsilon$ equations,
Eqns.~\ref{I1}-\ref{II4}, is described. Note that the functions
$v^{\epsilon}_1(y)$ and $v^{\epsilon}_2(y)$ are velocity fields and
satisfy Eqn.~\ref{directbc1}, the boundary conditions on $v$. The
quantities $q^{\epsilon}_1(y)$ and $q^{\epsilon}_2(y)$ are charge
densities. The functions ${\Psi}^{\epsilon}_1(y,z)$ and
${\Psi}^{\epsilon}_2(y,z)$ are electric potentials and satisfy
Eqns.~\ref{directbc2}-\ref{directbc5}, the boundary conditions on
$\Psi$.

To fully specify the order $\epsilon$ equations, we relate the $x$- and
$y$-components of the velocity field. This is accomplished by expanding the
incompressibility condition Eqn.~\ref{incomp} via the multiple-scales method.
At order $\epsilon^{1/2}$, Eqn.~\ref{incomp} is

\begin{equation}
\partial_x u_0(x,y) + \partial_y v_0(x,y) = 0 \;,
\end{equation}
which can be simplified using the order $\epsilon^{1/2}$ solution
for $v_0(x,y)$, Eqn.~\ref{orderhalfsol}, to

\begin{equation}
u_0(x,y) = A_0(X,T) \bar{u}_0(y) e^{i\kappa_0 x} + c.c. \;,
\label{unot}
\end{equation}
where
\begin{equation}
\bar{u}_0(y)=i\kappa_0^{-1} (\partial_y \bar{v}_0(y)) \;. \label{ubarnot}
\end{equation}

The solution of the first set of order $\epsilon$ equations,
Eqns.~\ref{I1}-\ref{I4}, begins with Eqn.~\ref{I2}. In the linear
stability calculation\cite{linear}, the functions
$\bar{v}_0(y)$ and $\bar{q}_0(y)$ are chosen to be even. So $\partial_y
\bar{v}_0(y)$ and $\partial_y \bar{q}_0(y)$ are odd. The derivative of the base
state charge density $Q(y)$ is even which means that if $v^{\epsilon}_1(y)$
is odd, then the product $Q(y)v^{\epsilon}_1(y)$ is also odd. Hence the
nonhomogeneous part of Eqn.~\ref{I2} can be expanded in a Fourier sine
series

\begin{equation}
  -(\partial_y \bar{v}_0(y)) \bar{q}_0(y)
	+ \bar{v}_0(y)(\partial_y \bar{q}_0(y))
  +Q(y)v^{\epsilon}_1(y) = \sum_{l=1}^{\infty} b_l sin(2\pi ly) \;.
\label{nonhomf3}
\end{equation}
The general solution of Eqn.~\ref{I2} is

\begin{eqnarray}
{\Psi}^{\epsilon}_1(y,0) & = & M sinh(2\kappa_0 y) - \sum_{l=1}^{\infty}
	[(2\pi l)^2 + (2\kappa_0)^2]^{-1} b_l sin(2\pi ly) \;. \label{generf3}
\end{eqnarray}
The coefficient $M$ is zero due to the boundary condition
${\Psi}^{\epsilon}_1(\frac{1}{2},0)=0$. Assuming a trial solution

\begin{equation}
v^{\epsilon}_1(y)=\sum_{m=1}^{N} E_m S_m(y) \;, \label{expansionf1}
\end{equation}
where $S_m$ is the odd Chandrasekhar function\cite{Chandrasekhar} with
$E_m=1$ for $m=1,...N$, yields the coefficients $b_l$ in Eqn.~\ref{nonhomf3}.
Equation~\ref{generf3} and the boundary conditions
Eqns.~\ref{directbc3}-\ref{directbc5} determine
${\Psi}^{\epsilon}_1(y,z)$ via the Helmholtz equation, Eqn.~\ref{I4}, which
is solved by a numerical relaxation method.
The function $q^{\epsilon}_1(y)$ is found by numerical
differentiation of ${\Psi}^{\epsilon}_1(y,z)$ in Eqn.~\ref{I3}. Then
$q^{\epsilon}_1(y)$ and ${\Psi}^{\epsilon}_1(y,0)$ are substituted
into Eqn.~\ref{I1} to calculate a new estimate of $v^{\epsilon}_1(y)$.
This process is repeated until $v^{\epsilon}_1(y)$,
$q^{\epsilon}_1(y)$, and ${\Psi}^{\epsilon}_1(y,z)$ are
self-consistently determined.

In the second set of order $\epsilon$ equations, Eqns.~\ref{II1}-\ref{II4},
Eqn.~\ref{II1} and the boundary conditions Eqn.~\ref{directbc1} indicate
that $v^{\epsilon}_2(y)=0$. The right-hand side of Eqn.~\ref{II2},
simplified via Eqn.~\ref{ubarnot}, is an odd function and
can be expanded in a Fourier sine series.
The general solution of Eqn.~\ref{II2} is

\begin{eqnarray}
{\Psi}^{\epsilon}_2(y,0) & = &
  - \sum_{l=1}^{\infty} (2\pi l)^{-2} a_l sin(2\pi ly) \;.
\label{generf6}
\end{eqnarray}
The variable ${\Psi}^{\epsilon}_2(y,z)$ is specified by solving
the Laplace equation, Eqn.~\ref{II4}, by a relaxation method, subject to
Eqn.~\ref{generf6} and the boundary conditions
Eqns.~\ref{directbc3}-\ref{directbc5}. The function $q^{\epsilon}_2(y)$ is
numerically calculated via Eqn.~\ref{II3}.

\section{}

The solution of the adjoint problem, Eqns.~\ref{adjequation}-\ref{adjbc5},
is discussed in this section. Substitution of the vector
$\bar{{\cal C}}_{b\kappa_0}$, Eqn.~\ref{adjsol}, into
Eqns.~\ref{adjequation1}-\ref{adjequation4} gives

\begin{eqnarray}
(\partial_y^2-\kappa_0^2)^2 \bar{v}_{b\kappa_0} -Q \bar{q}_{b\kappa_0} & = &
       0 \label{ADJ1} \;, \\
\kappa_0^2 {\cal R}_{c0} \bar{v}_{b\kappa_0} + \bar{\Psi}_{b\kappa_0}|_{z=0} &
= &
       0 \label{ADJ2} \;, \\
-\kappa_0^2 {\cal R}_{c0} Q
\bar{v}_{b\kappa_0}+(\partial_y^2-\kappa_0^2)\bar{q}_{b\kappa_0}
  +(\partial_z \bar{\Psi}_{b\kappa_0})|_{z=0^+}
  - (\partial_z \bar{\Psi}_{b\kappa_0})|_{z=0^-} & = & 0 \label{ADJ3} \;, \\
(\partial_y^2 +\partial_z^2 -\kappa_0^2) \bar{\Psi}_{b\kappa_0}
   & = & 0 \label{ADJ4} \;.
\end{eqnarray}
To simplify the notation, let $\bar{v}_{b0}=\bar{v}_{b\kappa_0}$,
$\bar{q}_{b0}=\bar{q}_{b\kappa_0}$, and
$\bar{\Psi}_{b0}=\bar{\Psi}_{b\kappa_0}$. The functions $\bar{v}_{b0}(y)$,
$\bar{q}_{b0}(y)$, and $\bar{\Psi}_{b0}(y,z)$ are expanded as

\begin{eqnarray}
\bar{v}_{b0}(y) & = & \sum_{m=1}^{\infty} B_m
       \bar{v}_{b0 m}(y) \;, \label{adjsumsol1} \\
\bar{q}_{b0}(y) & = & \sum_{m=1}^{\infty} B_m
       \bar{q}_{b0 m}(y) \;, \label{adjsumsol2} \\
\bar{\Psi}_{b0}(y,z) & = &
      \sum_{m=1}^{\infty} B_m \bar{\Psi}_{b0 m}(y,z) \;,
\label{adjsumsol3}
\end{eqnarray}
where $\bar{v}_{b0 m}(y)$, $\bar{q}_{b0 m}(y)$, and
$\bar{\Psi}_{b0 m}(y,z)$ satisfy the adjoint boundary conditions,
Eqns.~\ref{adjbc1}-\ref{adjbc5}.

With the solutions Eqns.~\ref{adjsumsol1}-\ref{adjsumsol3},
Eqn.~\ref{ADJ1} implies that

\begin{eqnarray}
(\partial_y^2-\kappa_0^2)^2 \bar{v}_{b0 m}(y) -Q(y) \bar{q}_{b0 m}(y)
  & = & 0 \;.  \label{eqvbm1}
\end{eqnarray}
Since $\bar{q}_{b0 m}(y)$ must satisfy
$\bar{q}_{b0 m}(y=\pm \frac{1}{2})=0$, let
$\bar{q}_{b0 m}(y)= cos((2m-1)\pi y)$.
The product of $Q(y)$ and $cos((2m-1)\pi y)$ is even and can be
represented by a Fourier cosine series. The general solution of
Eqn.~\ref{eqvbm1} is

\begin{eqnarray}
\bar{v}_{b0 m}(y) & = &
   M_1 cosh(\kappa_0 y) + M_2 y sinh(\kappa_0 y) +
 \sum_{l=0}^{\infty} [(2\pi l)^2 + \kappa_0^2]^{-2} b_{ml} cos(2\pi ly) \;,
\label{vbarb0m}
\end{eqnarray}
where the constants $M_1$ and $M_2$ are specified by the boundary
conditions $\bar{v}_{b0 m}(y=\pm \frac{1}{2})=
(\partial_y \bar{v}_{b0 m})(y=\pm \frac{1}{2})= 0$ to be

\begin{eqnarray}
M_1 & = & -2 (\kappa_0 + sinh(\kappa_0))^{-1}
[ sinh(\kappa_0 / 2) + (\kappa_0 / 2)cosh(\kappa_0 / 2) ]
\sum_{l=0}^{\infty} (-1)^l [(2\pi l)^2 + \kappa_0^2]^{-2} b_{ml}
  \nonumber \;, \\
M_2 & = & -2 (\kappa_0 +sinh(\kappa_0))^{-1}
[ -\kappa_0 sinh(\kappa_0 / 2) ]  \sum_{l=0}^{\infty} (-1)^{l} [(2\pi l)^2
+\kappa_0^2]^{-2} b_{ml} \nonumber \;.
\end{eqnarray}

Substitution of Eqns.~\ref{adjsumsol1}-\ref{adjsumsol3} into Eqn.~\ref{ADJ2}
and Eqn.~\ref{ADJ4} yields

\begin{equation}
\bar{\Psi}_{b0 m}(y,0)=-\kappa_0^2 {\cal R}_{c0} \bar{v}_{b0 m}(y) \;,
\label{adjpsiy0}
\end{equation}
and
\begin{equation}
(\partial_y^2 + \partial_z^2 - \kappa_0^2)\bar{\Psi}_{b0 m}(y,z) = 0 \;.
\label{adjpsiyz0}
\end{equation}
The latter is a Helmholtz equation subject to the adjoint boundary
conditions Eqns.~\ref{adjbc4} and \ref{adjbc5}, with
$\bar{\Psi}_{b0 m}(y,0)$, $|y| \leq \frac{1}{2}$, given by
Eqn.~\ref{adjpsiy0}. Equation \ref{adjpsiyz0} is solved by a
numerical relaxation method.

Using Eqns.~\ref{adjsumsol1}-\ref{adjsumsol3} to expand Eqn.~\ref{ADJ3}
leads to

\begin{eqnarray}
\sum_m B_m [ -\kappa_0^2 {\cal R}_{c0} Q\bar{v}_{b0 m}
   +(\partial_y^2-\kappa_0^2)\bar{q}_{b0 m}
  +(\partial_z \bar{\Psi}_{b0 m})|_{z=0^+}
  -(\partial_z \bar{\Psi}_{b0 m})|_{z=0^-} ] & = & 0 \;.
\label{secular1}
\end{eqnarray}
Since $\bar{v}_{b0 m}(y)$, $\bar{q}_{b0 m}(y)$, and
$\bar{\Psi}_{b0 m}(y,z)$ are known functions, Eqn.~\ref{secular1}
implies that the coefficients $B_m$ vanish unless the compatibility
condition
\begin{equation}
\Biggl{\|} T_{nm} \Biggr{\|} = \Biggl{\|} \int_{-1/2}^{1/2} dy
 [ -\kappa_0^2 {\cal R}_{c0} Q\bar{q}_{b0 n}\bar{v}_{b0 m}
  +\bar{q}_{b0 n}(\partial_y^2-\kappa_0^2)\bar{q}_{b0 m}
  +2\bar{q}_{b0 n} (\partial_z \bar{\Psi}_{b0 m})|_{z=0^+} ]
\Biggr{\|}
   =  0 \;,
\label{secular2}
\end{equation}
is satisfied. Note that the relation
$(\partial_z \bar{\Psi}_{b0 m})|_{z=0^-} =
 -(\partial_z \bar{\Psi}_{b0 m})|_{z=0^+}$, which is analogous to the
discontinuity in the electric field $\partial_z \bar{\Psi}_{0 m}$ across
the two surfaces of the film in the direct problem, is used to derive
Eqn.~\ref{secular2}. In Eqn.~\ref{secular2}, the values of $\kappa_0$ and
${\cal R}_{c0}$ are fixed to be those obtained from the linear stability
analysis. The coefficients $B_m$ are determined by the matrix equation
\begin{equation}
T_{nm}B_m=0 \;, \label{secular3}
\end{equation}
where $T_{nm}$ is given in Eqn.~\ref{secular2}.
These coefficients are then substituted into
Eqns.~\ref{adjsumsol1}-\ref{adjsumsol3} to generate
$\bar{v}_{b0}(y)$, $\bar{q}_{b0}(y)$, and
$\bar{\Psi}_{b0}(y,z)$. This specifies $\bar{{\cal C}}_{b\kappa_0}$
and the general solution vector ${\cal C}_b$,
Eqn.~\ref{adjsol}.


\begin{table}
\caption{Numerical results}
\label{coefftable}
\begin{tabular}{lcc}
   & Nonlinear Analysis & Linear Stability \\ \hline

Wire Electrode Geometry &  &   \\
\hline
critical wavenumber, ${\kappa}_0$ & 4.7467 & 4.744 \\
critical control parameter, ${\cal R}_{c0}$ & 76.855 & 76.77 \\
correlation length, ${\xi}_0$ & 0.28484 & 0.2843 \\
time scale, $\tau_0$ & 0.35072 & 0.351 \\
nonlinear coupling, $g_0$ & 1.74602 & - \\
\\
\hline
\hline
Plate Electrode geometry & & \\
\hline
critical wavenumber, ${\kappa}_0$ & 4.2239 & 4.223 \\
critical control parameter, ${\cal R}_{c0}$ & 91.855 & 91.84 \\
correlation length, ${\xi}_0$ & 0.29743 & 0.2975 \\
time scale, $\tau_0$ & 0.37155 & 0.372 \\
nonlinear coupling, $g_0$ & 2.8424 & - \\
\end{tabular}
\end{table}

\vfill\eject

\end{document}